\documentclass[aps,prb,twocolumn,showpacs,floatfix]{revtex4}
\usepackage{dcolumn}
\usepackage{bm}
\usepackage{graphicx}
\usepackage{bm}

\begin{document}

\title{Magnetic glass in Shape Memory Alloy : Ni$_{45}$Co$_{5}$Mn$_{38}$Sn$_{12}$}

\author{Archana Lakhani}
\email{archnalakhani@csr.res.in}\affiliation{UGC-DAE Consortium for Scientific Research, University Campus, Khandwa Road, Indore-452001,India}

\author{A. Banerjee}
\affiliation{UGC-DAE Consortium for Scientific Research, University Campus, Khandwa Road, Indore-452001,India}

\author{P. Chaddah}
\affiliation{UGC-DAE Consortium for Scientific Research, University Campus, Khandwa Road, Indore-452001,India}

\author{X. Chen}
\affiliation{School of Materials Science and Engineering, Nanyang Technological University, N4.1-01-18, 50 Nanyang Avenue, Singapore -639798}

\author{R. V. Ramanujan}
\affiliation{School of Materials Science and Engineering, Nanyang Technological University, N4.1-01-18, 50 Nanyang Avenue, Singapore -639798}

\begin{abstract}
The first order martensitic transition in the ferromagnetic shape memory alloy Ni$_{45}$Co$_{5}$Mn$_{38}$Sn$_{12}$ is also a magnetic transition and has a large field induced effect. While cooling in the presence of field this first order magnetic martensite transition is kinetically arrested. Depending on the cooling field, a fraction of the arrested ferromagnetic austenite phase persists down to the lowest temperature as a magnetic glassy state, similar to the one observed in various intermetallic alloys and in half doped manganites. A detailed investigation of this first order ferromagnetic austenite (FM-A) to low magnetization martensite (LM-M) state transition as a function of temperature and field has been carried out by magnetization measurements. Extensive cooling and heating in unequal field (CHUF) measurements and a novel field cooled protocol for isothermal MH measurements (FC-MH) are utilized to investigate the glass like arrested states and show a reverse martensite transition. Finally, we determine a field -temperature  (HT) phase diagram of Ni$_{45}$Co$_{5}$Mn$_{38}$Sn$_{12}$ from various magnetization measurements which brings out the regions where thermodynamic and metastable states co-exist in the HT space clearly depicting this system as a 'Magnetic Glass'.  

\end{abstract}



\maketitle\

\section{\textbf{INTRODUCTION}}
The kinetics of any first order phase transition (FOPT) is governed by time required for extracting the latent heat of the system and can be arrested by rapid enough cooling \cite{1}. When a magnetic first order transition is kinetically arrested it gives rise to the states which are magnetically ordered metastable states and are referred as 'magnetic glasses' \cite{2}. 

Kinetic arrest of martensite transition (MT) in ferromagnetic shape memory alloys (FSMAs) has been reported extensively in recent years, especially in NiMnIn and NiMnSn based FSMAs [3-22]. Apart from the technological importance of FSMAs, they are also very good candidates for the fundamental studies in field -temperature (HT) phase space since first order martensite transition is influenced by both field and temperature. The parent austenite phase shows a stronger ferromagnetism than that of the martensite phase and at the martensitic transition temperature (T$_m$), a sharp change in magnetization is observed in thermo-magnetization measurements. T$_m$ decreases on increasing the magnetic field of measurement. The kinetics associated with magnetic FOPT gets hindered in many of these materials resulting in phase coexisting states at low temperature which comprises of an arrested (metastable) state and the transformed (stable) state. In spite of the wide ranging studies justifying the presence of kinetic arrest of martensite transformation in FSMAs, the magnetic glass concept has been discussed only in a few In and Sn based FSMAs [18-22]. 

In general, magnetic field may enhance or reduce the kinetic arrest, depending on whether the low temperature equilibrium state has lower or higher value of magnetization \cite{23,24}. It has been noted that by cooling in a certain field (H$_c$) and then heating in a different field (H$_w$), this glass like arrested state (GLAS) can be de-arrested and on further heating a reverse magnetic transition is observed. Hence by cooling and heating in unequal fields (CHUF), a re-entrant transition can be seen but for only the appropriate sign of (H$_c$-H$_w$) \cite{24}. In the case of FSMAs, since the high temperature phase is ferromagnetic austenite (FM-A) while the low temperature phase is low magnetization martensite (LM-M), this reentrant transition will be seen only for the positive sign of (H$_c$-H$_w$). Kainuma's group has reported magnetization measurements on cooling in various fields and warming in lower fields (0.05T) on Ni$_{50}$Mn$_{34}$In$_{16}$ alloy and observed the `unfreezing of P+M coexisting state' \cite{14}. This `dearrest' stated as `unfreezing' is analogous to `devitrification' observed in manganites with charge ordered ground state  [25-27].  These results are consistent with what is expected in a magnetic glass; however the well established CHUF protocol provides a crucial test for validating whether the kinetically arrested state corresponds to a magnetic glass. The CHUF protocol requires that the measurements are done in the fields both below and above H$_c$, i.e. for both positive and negative values of (H$_c$-H$_w$). 

In this communication our studies on Ni$_{45}$Co$_{5}$Mn$_{38}$Sn$_{12}$ ribbon sample demonstrate all the features crucial to unambiguously prove the magnetic glass state by various thermo-magnetization measurements, including CHUF measurements. Besides, we highlight an important aspect that the transition between LM-M and FM-A phases can not be achieved at low temperature by the conventional protocol of isothermal field variation of the ZFC state because of the limit of experimentally accessible field range. By following the novel  protocol of field cooling and then isothermally reducing the field we are able to observe the transition between LM-M and FM-A phases within the accessible field range even at the lowest temperature. We justify this by constructing the qualitative phase diagram identifying bands of supercooling (SC), superheating (SH) and kinetic arrest (KA) from the variety of measurements spanning the HT space and check its self consistency. As has been discussed in detail in reference[23] the SC limit T*(H), the SH limit T**(H) and the kinetic arrest line T$_K$(H) are broadened into bands in system with quenched disorder. Thus we show that the technologically useful FSMAs are magnetic glasses, as shown earlier in manganites.

\section{\textbf{EXPERIMENTAL DETAILS}}

The sample Ni$_{45}$Co$_{5}$Mn$_{38}$Sn$_{12}$ (Sn12) used in this study is prepared by arc melting the high purity elements into buttons and preparing ribbons from these buttons by melt spinning in an inert atmosphere. The composition and crystal structure were determined by an energy dispersive X-ray spectrometer (EDXS) attached to a scanning electron microscope (SEM), and powder X-ray diffraction (XRD).  Details of sample preparation and characterization can be seen in reference 20. The DC-Magnetization was measured with a commercial 14 Tesla PPMS- VSM (M/s: Quantum Design).

\section{\textbf{RESULTS AND DISCUSSIONS}} 	

Sn12 sample exhibits a clear first order phase transition with respect to field and temperature, displaying broad hysteresis. The high magnetization ferromagnetic austenite (FM-A) phase transforms to the low magnetization martensite (LM-M) phase and \emph{vice versa} on cooling and heating, respectively \cite{20,21}. The first order nature of the martensite transition with respect to field in this sample was established in our earlier work \cite{20} by performing zero field cooled Magnetization vs Field (MH) measurements during cooling and heating \cite{28}. The MH isotherms in both the cases (cooling and heating) were completely different for (T*$\sim$75K)$<$T$< $(T**$\sim$~215K). Similarly, in the plot of Magnetization v/s Temperature i.e. M(T), a broad hystersis is observed in the same temperature range during cooling and heating. These results signify that the fractions of coexisting metastable martensite and austenite phases not only vary with field and temperature but also depend on the path followed in the HT space. These metastable states at low temperature are governed by an interplay of supercooled and kinetically arrested states, because neither the supercooled nor arrested state is the equilibrium state of the system. \cite{29}

	Figure 1 shows magnetization as a function of temperature in three different measurement protocols viz: zero field cooled warming (ZFC), field cooled cooling (FCC) and field cooled warming (FCW) at 1, 3, 5 and 9T. The FCC and FCW curves shown here were reported in reference 20 and 21. They are reproduced here in a replotted version along with the ZFC curve to explain the thermomagnetic irreversibility in these systems at low temperatures. As shown in figure 1(a) the cooling and heating curves at 1T exhibit a clear first order phase transition from FM-A phase to LM-M phase, showing a broad hysteresis in the range of (T*$\sim$75K)$<$T$<$(T**$\sim$215K). At higher fields the MT gets gradually inhibited and completely arrested at 9T, as shown in figures 1(b-d). As a result, various co-existing metastable states are observed, depending on the magnitude of the magnetic field during cooling and heating. These `kinetically arrested states' are due to slowing down of growth of the martensite phase from the supercooled austenite phase, similar to the viscous retardation in structural glasses. In structural glasses the liquid to crystal freezing is arrested, while in a magnetic glass the magnetic re-ordering is arrested. In both cases the material is cooled through the first order transition temperature without extracting latent heat. While in a jamming process translational kinetics is arrested; in the formation of both these glasses the specific heat is extracted, while the latent heat (which has a different coupling with thermal conduction process) can not be extracted \cite{1}. In both cases the higher entropy phase persists down to the lowest temperature \cite {30}. In this sense the magnetic glass and the structural glass address similar physics. However in structural glasses pressure is the variable parameter. The experimental advantage of magnetic field (H) over pressure (P) is that it can be easily varied reversibly without a control medium, and its variation does not interfere with temperature control.

\begin{figure}[t]
	\begin{center}
	\includegraphics[width=0.8\columnwidth]{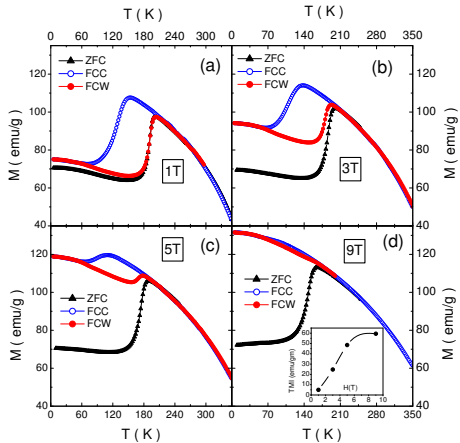}
	\end{center}
	\caption{(Color online)(a-d) M vs T plots for Sn-12 sample obtained in ZFC, FCC and FCW mode in applied fields of H = 1, 3, 5 and 9T respectively.  FCC and FCW data are taken from ref. 20 and 21. }
	\label{Figure 1}
\end{figure}

As the first order MT gets arrested in presence of field, the low temperature states turn out to be non-ergodic. At 1T, there is no significant difference between ZFC and FCW curves as shown in figure 1(a). However, at field above 1T, the hysteresis width $\bigtriangleup$T does not change significantly but the area enclosed by the hysteresis reduces and the difference between M$_{ZFC}$ and M$_{FCW}$ at low temperature increases. This difference is known as thermomagnetic irreversibility (TMI), which is defined as TMI= M$_{FCW}$-M$_{ZFC}$ at 5K. The TMI at low temperatures could be due to spin glass behavior, pinning of variants, hindrance of domain motion or glass like arrest of dynamics. Nevertheless for the first three cases, this irreversibility decreases with increasing field; hence the ZFC and FCW magnetization curves would merge, in contrast to the present case.  Here, TMI increases with increase in field as shown in the inset of figure 1(d). This rise in thermomagnetic irreversibility is due to the kinetic arrest of the martensite phase transition at the low temperatures which gives rise to non-ergodic states having coexisting FM-A and LM-M phases. We have almost 100\%  LM-M phase fraction when cooled in zero field but on increasing the cooling field the ferromagnetic austenite fraction increases. We have estimated this phase fraction by assuming the field cooled magnetization values at 5K and 9T as 100\% FM-A as there is no FM-A to LM-M transition on heating and cooling at this field \cite{21}. The TMI at 5K gives an amount of the fraction of arrested FM-A mixed with the LM-M phase. This attribute classifies this system as a magnetic glass. 

This system undergoes a complete transformation below 1T (H$_1$), while above 8T (H$_2$) the martensite transformation is almost hindered resulting in a completely austenite phase at low temperature. However at field values between H$_1$ and H$_2$, we achieve a fraction of transformed martensite phase along with an arrested austenite phase at low temperatures \cite{21}.  In the earlier studies of manganites and intermetallic alloys [25-28,33], referred as magnetic glasses, the arrest of kinetics across the first order phase transformation is established from the fact that the virgin MH curve lies outside the envelope curve in the isothermal magnetization curves at lowest temperatures. In these cases, the samples have undergone a first order transformation within the available field and temperature range and the observed low temperature anomaly is inferred as the arrested kinetics of the first order transition. In all these cases, the virgin state considered are the zero field cooled state from well above the superheating limit of the samples. They have also shown a non monotonic behavior of metamagnetic transition field with respect to temperature in zero field cooled MH isotherms, which is an important experimental method of studying the arrested kinetics in these samples. 

We performed similar zero field cooled isothermal magnetization (ZFC-MH) experiments in the range of 5-150K, as shown by the symbols (green) in figure 2(a-h). The sample is cooled from temperatures well above MT temperature. After cooling in zero field, the low temperature state of the sample is in a low magnetization martensite (LM-M) phase, as seen by dashed lines in figure 2(a). The initial field increasing cycle at 5, 15 and 25K shows a sharp increase in magnetization up to $\sim$1T and then almost saturates, while the return cycle follows the same path indicating the soft ferromagnetic behavior. However, we did not observe the reverse martensite transition (RMT) at low temperatures up to 14T. For temperatures above 75K, the rise in magnetization in the forward field cycle followed by an opening of hysteresis indicates the onset of the martensite to austenite phase transformation, which confirms the RMT in the temperature range of 75-150K. Therefore to investigate the absence of such RMT  at low temperature, which we attribute to kinetic arrest in this system, we designed a novel protocol of field cooled MH isotherms (FC-MH), as shown by the solid lines (red) in figure 2(a-h). For FC-MH measurements, we cooled the sample in 8T from 350K, which is well above the superheating limit ($\sim$210K) of the sample, and considered the field cooled state as the initial state of the sample. On cooling in 8T, the sample acquires maximum fraction of FM-A phase as an arrested phase. In each case, the sample is cooled in 8T to the respective temperatures, and then magnetization is measured while isothermally raising the field up to 12T, subsequently cycling from $12T \rightarrow{0}\rightarrow{(-12T)}\rightarrow{0T}$. 

We explain these measurements by the schematic band diagram specified for magnetic glasses shown in figure 2(i). Since FOPT (here MT) induced by field and temperature is broadened in HT space, as a result the sharp lines corresponding to the transition temperature (here T$_m$) as well as the spinodal lines of supercooling (H*T*) and superheating (H**, T**) broaden into bands \cite{29,31,32}.  As discussed earlier \cite{24,25,28,33}, if kinetic arrest occurs below (H$_k$,T$_k$) line in a pure system, the disordered system would have an (H$_k$,T$_k$) band. The kinetic arrest band in the HT space contains a set of lines below which the dynamics of MT is hindered on the time scales of the experiments, as in structural glasses. These lines in the band would correspond to a local region of the sample and therefore a correlation between the positions of (H$_k$,T$_k$) and (H*,T*) bands can be established. An anticorrelation between the supercooled and kinetically arrested regions has been proposed and experimentally confirmed in many materials termed as magnetic glasses having first order magnetic transitions [23-28,33,34]. 

On reducing the field from point A(5K,12T), approached by following the path XA, to point D(5K,0T) as shown in the HT phase diagram in figure 2(i), we observe a step at 1.5T in MH curve (figure 2(a)), which is expected to be due to the de-arrest of arrested high temperature phase as it crosses the KA band \cite{23}. At higher temperatures, e.g., 15K, a signature of dearrest of FM-A phase at the field marked by a circle in figure 2(b) is observed by a clear slope increase at dearrest field H$_d$. This field H$_d$ marked by circles in figure 2 increases on increasing the temperature and the dearrested fraction also increases upto 75K. At temperatures greater than 75K, $\delta$M (estimated by the drop that occurs with this increased slope) decreases and we observe smaller fraction of the dearrested phase. The corresponding arrested FM-A phase is reduced due to the presence of higher supercooled fraction at temperatures between T= 75-150K. Hence, the width of the linear slope in the return field cycle increases from 5K-75K and then decreases up to 150K. Considering T=150K at point C in the schematic diagram 2(i), if we traverse on the field axis, we observe a completely symmetrical and reversible MH curve as shown in figure 2(h). At this temperature whether we approach by zero field cooling (from point F) or cooling in a finite field (from point C), one would get the same behavior, since 150K lies outside the (H$_k$,T$_k$) band.

 	\begin{figure}[b]
	\begin{center}
	\includegraphics[width=0.58\columnwidth]{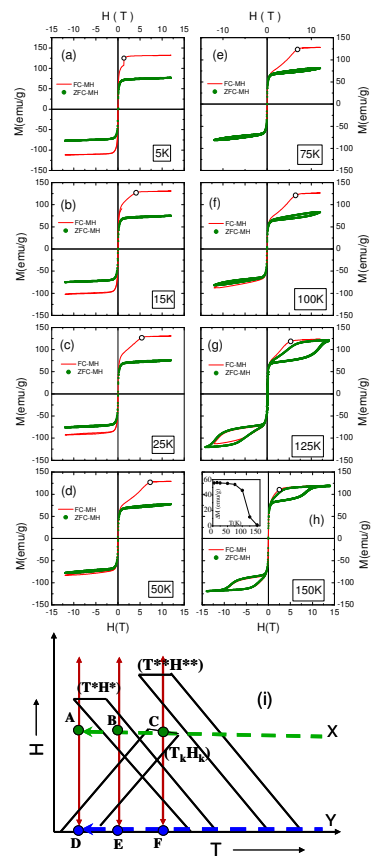}
	\end{center}
	\caption{ (Color online) (a-h):Isothermal magnetization (solid red lines) at various temperatures from 5K to 150K after cooling the sample in 8 Tesla from 350K to the respective temperatures after raising the field to 12T. For comparison the zero field cooled M vs H curves (green symbols) are shown along with the field cooled data. Inset (2h): Difference in magnetization for FC-MH and ZFC-MH at 8T for various temperatures obtained from figure 2(a-h); (i): Schematic diagram showing the path of 8T field cooled (green, X) and zero field cooled (blue, Y) MH isotherms.}
	\label{Figure 2}
	\end{figure}	
	Since ZFC -MH shows a fully martensite state and FC-MH shows fully arrested metastable state at 5K, we can estimate the arrested FM-A fraction from the difference in magnetization ($\bigtriangleup$M) between FC-MH and ZFC-MH at 8T for various temperatures as shown in the inset of Fig 2(h). After cooling in 8T field, the arrested FM-A fraction is almost constant below 75K, which is T* at 8T for this sample. Thereafter, due to the presence of supercooled austenite fraction $\bigtriangleup$M gradually decreases and almost vanishes on heating above 150K. At T$\sim$150K the kinetic arrest and the supercooled band tend to overlap. Hence, FC-MH measurements demonstrate the de-arrest of the glass like arrested metastable FM-A states below 150K.

Cooling and heating in unequal field (CHUF) measurements confirm the kinetically arrested metastable states at low temperatures by observing the de-arrest of arrested phase \cite{34,35}. They are also used to distinguish the equilibrium phase from glass like arrested co-existing phases \cite{33,35}.  To verify GLAS at low temperatures by observing the de-arrest on warming, we have carried out four sets of CHUF measurements. In the first two set of measurements, magnetization is performed by cooling in a constant cooling field (H$_c$), in various different measuring fields (H$_w$) for both H{$_c$}{$<$}H{$_w$} and H{$_c$}{$>$}H{$_w$}. In the other two sets, the sample is cooled in various different fields and the measurement field is kept constant during heating. The CHUF set is said to be complete when the measurements are performed in the field below and above the cooling and/or warming field. To confirm the re-entrant transition mentioned above for demonstrating the glass like arrested state, a complete set of CHUF measurements is required.

In the first two sets of CHUF measurements, the constant cooling fields (H$_c$) of 6T and 3T is chosen, for which one observes a finite fraction of both arrested and transformed phases at low temperature (figure 3(a) and (b)). In each case the sample is cooled from 350K down to 10K. At 10 K, the field is reduced isothermally to the respective measuring field (H$_w$) and measurement is done while heating. For the other two set of measurements (figure 3(c)and(d)), the sample is cooled in various fields (H$_c$) ranging from 0T to 8T down to 10K and then the field is isothermally increased or decreased to the constant measuring field (H$_w$) of 4T and 1T, M(T) is measured while heating. 
	
 	\begin{figure}[t]
	\begin{center}
	\includegraphics[width=0.8\columnwidth]{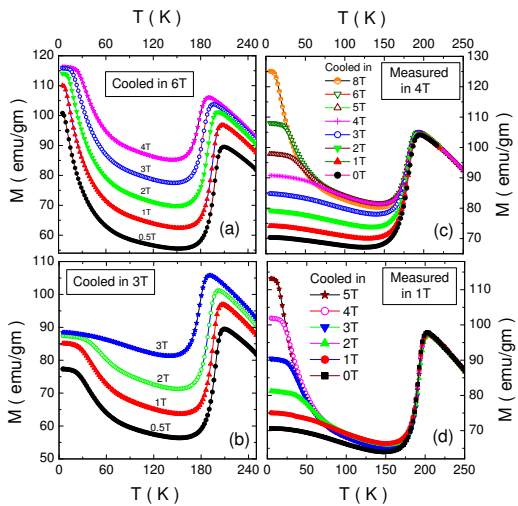}
	\end{center}
	\caption{ (Color online) Magnetization as a function of temperature using CHUF (cooling and heating in unequal field) protocol. (a,b) Sample is cooled in a constant field of 6 and 3T and measurements are carried out in various different fields. (c,d ) Sample is cooled under different magnetic fields whereas measurements during warming are carried out in 4 and 1 T respectively. Fig 4(c) is reproduced from ref. 20.}
	\label{Figure 3}
	\end{figure} 
	
As shown by M(T) in figure 3(a) when system is cooled in 6T and heated in 0.5, 1, 2, 3, and 4T an initial drop is observed in magnetization and this drop is more profound when the difference between H$_c$ and H$_w$ is large. Similar trend is observed in fig 3(b) when system is cooled in 3T and warmed in 0.5, 1 and 2T, except that the initial arrested fraction is smaller.  The fall in magnetization on increasing the temperature indicates the de-arrest of FM-A to LM-M state, i.e., the arrested FM-A phase (untransformed phase during cooling in 6T/3T) transforms to the LM-M phase on heating in lower fields. On the other hand, if the warming field is kept constant while the sample was cooled in different fields, a clear drop in magnetization in figures 3(c and d) indicates the transition to LM-M state when H{$_c$}{$>$}H{$_w$} i.e., for the positive values of (H$_c$-H$_w$). Only one transition back to the FM-A state is observed at higher temperature while warming in 4T after cooling in 0, 1, 2 and 3T as well as during warming in 1T after cooling in 0T. This also shows that the LM-M is an equilibrium state and FM-A state is a glass like arrested state (GLAS). 

For the systems, where T$_m$ (and T*) falls with increase in magnetic field, H{$_c$}$>$H{$_w$} is required for observing the de-arrest; T* would be lower during cooling in H$_c$ and higher during warming in H$_w$. Such a transformation to the low temperature equilibrium phase caused by thermal energy is analogous to the phenomenon of devitrification of the glassy phase, which has been demonstrated in half doped manganites and in many intermetallic alloys classified as magnetic glasses.  Devitrification of glasses has been used as evidence for metallic glasses \cite{36}. Another transition of LM-M phase back to FM-A phase at higher temperatures is analogous to the melting of devitrified crystalline equilibrium phase on increasing temperature. Both these transitions are seen only when the difference between cooling field and warming field is positive. Hence, the de-arrest of FM-A to LM-M is observed when H{$_c$}$>$H{$_w$} and when the difference between cooling and heating field (H$_c$ - H$_w$) decreases, dearrest starts at higher temperatures. Hence, regions which have higher T* will have lower T$_k$. This relation is universal for all magnetic glasses \cite{23,37}. 

Based on the results from all the magnetization measurements, we have developed a quantitative phase diagram shown in figure 4, similar to the one given for the magnetic glasses mentioned above. The filled squares represent the supercooling limit for the corresponding fields obtained from the peak points of FCC magnetization data and empty squares are taken from the points of slope change in the return envelope curve of ZFC- MH measurements. The triangular symbols represent the superheating limit for various fields obtained from different measurements, as indicated in the legends of figure 4 and explained in the figure caption. The filled and empty circles correspond to the limit of kinetic arrest acquired from CHUF magnetization measurements performed for H$_c$ = 6T and 3T respectively. Star symbols represent the upper limit of kinetic arrest band obtained from FC-MH isotherms. We observe that the temperature and fields corresponding to the superheating and supercooling limits obtained from diverse measurements like FCW-MT, ZFC-MH, ZFC-MT, 3/6T cooled CHUF and FCC-MT, fall uniformly in the respective bands. Similarly, field and temperatures corresponding to the kinetic arrest edges obtained by quite different sets of measurements mentioned above, are consistent. Hence, validity of our phenomenological model is reinforced by the results obtained from the various measurement protocols falling in the same bands.
This phase diagram gives the quantitative information about the phase fraction kinetically arrested across the H$_k$T$_k$ band on cooling and/or applying the field as well as about the de-arrested fraction on warming and/or reducing the field.  This phase diagram also illustrates that, although the metamagnetic transition from FM-A to LM-M state at the lowest temperature takes place at $\sim$30T for this sample, which is far above the available field range of measurements, we can visualize these features at much lower fields in the field reducing cycle of FC-MH by observing the de-arrest of FM-A state to LM-M state. 

 	\begin{figure}[t]
	\begin{center}
	\includegraphics[width=0.7\columnwidth]{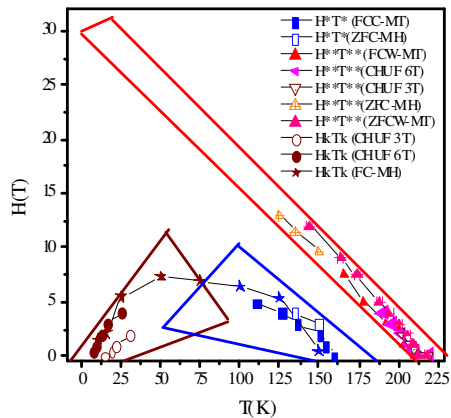}
	\end{center}
	\caption{ (Color online) Schematic H-T Phase diagram showing supercooling (Blue), superheating (red) and kinetic arrest bands (wine) obtained from various measurement protocols for Ni$_{45}$Co$_{5}$Mn$_{38}$Sn$_{12}$ ribbon. FCC-MT: Magnetization w.r.t temperature during field cooled cooling, FCW-MT: Magnetization vs temperature during field cooled warming, ZFC-MT: Magnetization w.r.t temperature after zero field cooling. ZFC-MH: Isothermal magnetization vs field after zero field cooling. CHUF-3T/6T: M w.r.t T in various fields after cooling in 3T/6T. FC-MH: Isothermal magnetization w.r.t field after cooling in 8T. Peak points of M in the MT curves and points of slope change in MH isotherms are taken as the representative limiting points for SC, SH and KA bands}
	\label{Figure 4}
	\end{figure} 
The formation of magnetic glass in Ni$_{50}$Mn$_{50-x}$Sn$_{x}$ FSMAs could be related to the mechanism of structurally modulated phases observed experimentally below MT. The true structural ground state in these alloys is found to be experimentally ambiguous at low temperatures as various modulated states are observed which are metastable and vary with x concentration \cite{38,39}. Therefore as the degeneracy in structure of FSMAs is affected by small variation of Sn/Mn concentration, various metastable magnetic states can be formed by the kinetic arrest of MT in varying magnetic fields, leading to the formation of magnetic glass.      

\section{\textbf{CONCLUSION}}

We have performed a comprehensive study of field induced first order FM-A to LM-M transition with respect to temperature and field by variety of magnetization measurements on the ferromagnetic shape memory alloy ribbons of Ni$_{45}$Co$_{5}$Mn$_{38}$Sn$_{12}$. The kinetically arrested glass like metastable states are revealed by cooling and/or applying the field and their de-arrest is observed on warming and/or reducing the field by various thermomagnetic measurements. This phenomenon is validated by a CHUF protocol to study the arrested high temperature phase and observe the re-entrant transition for positive values of H$_c$-H$_w$. The co-existing states have been produced and higher arrested fraction at higher cooling fields has been confirmed by the novel protocol of `field cooled MH isotherms' where the 8T field cooled state is considered as the initial state of the system.  Based on these measurements we regard this ferromagnetic shape memory alloy system as a `Magnetic Glass', similar to charge ordered ground state manganites.  

\section{\textbf{ACKNOWLEDGMENT}}
We thank S. Dash for help in magnetization measurements. RVR and CX acknowledge support from National Research Foundation, Singapore through the CREATE program on Nanomaterials for Energy and Water Management.


\begin{thebibliography}{39}
\bibitem[1] {1} P. Chaddah and A. Banerjee, arXiv:1004.3116v2, (2012) and references therein.
\bibitem[2] {2}	M. K. Chattopadhyay, S. B. Roy and P. Chaddah, Phys. Rev.B  72, (2005) 180401(R) .
\bibitem[3] {3}	J.I. Perez-Landazabal, V. Recarte, V. Sanchez-Alarcos, S. Kustov and E. Cesari , (In press) J. of Alloys and Compounds (2012)  
\bibitem[4] {4}	X. Xu, W. Ito, I. Katakura, M. Tokunaga, R. Kainuma, Scripta materialia, 65 (2011) 946
\bibitem[5] {5}	J. I. Perez-Landazabal, V. Recarte and V. Sanchez-Alarcos, C. Gomez-Polo, S. Kustov and E. Cesari, J. of Appl. Phys. 109, (2011), 093515
\bibitem[6] {6}	Yong-hee Lee, Mitsuharu Todai, Takahiro Okuyama, Takashi Fukuda, Tomoyuki Kakeshita and Ryosuke Kainuma, Scripta Materialia, 64, (2011) 927
\bibitem[7] {7} Zhigang Wu, Zhuhong Liu, Hong Yang, Yinong Liu and Guangheng Wu, Appl. Phys. Lett,  98 (2011) 061904
\bibitem[8] {8}	S. Kustov, I. Golovin, M. L. Corro and E. Cesari, J. of Appl. Phys. , 107, (2010) 053525.
\bibitem[9] {9}	J. L. Sanchez Llamazares, B. Hernando, J. J. Sunol, C. Garcia, C. A. Ross, J. of Appl. Phys. 107, (2010) 09A956
\bibitem[10] {10}	V. K. Sharma, J. D. Moore, M. K. Chattopadhyay, Kelly Morrison, L. F. Cohen, S. B. Roy, J. of Phys. Cond.  Mat., 22,(2010) 486007
\bibitem[11] {11}	 R. Y. Umetsu, Y Kusakari, T Kanomata, K Suga, Y Sawai, K Kindo, K Oikawa, R Kainuma and K Ishida, J. Phys D : Appl. Physics, 42 (2009) 075003.
\bibitem[12] {12} S. Chatterjee, S. Giri, S. Majumdar and S.K. De, Phys. Rev. B 77, (2008) 224440 
\bibitem[13] {13}	W. Ito, R.Y. Umetsu, R. Kainuma, T. Kakeshita, K. Ishida, Scripta Materialia 63, (2010) 73
\bibitem[14] {14}	R. Y. Umetsu, W. Ito, K. Ito, K. Koyama, A. Fujita, K. Oikawa , T. Kanomata, R. Kainuma and K.Ishida, Scrip. Mater., 60 (2009) 25-28
\bibitem[15] {15}	M. Wang, L.Wang , Y. Liu, B.C. Zhao, Y. Yang abd H. Zhang, J. Appl. Phys. 106, (2009) 063909
\bibitem[16] {16}	W. Ito, K. Ito, R. Y. Umetsu, R. Kainuma, K. Koyama, K. Watanabe, A. Fujita, K. Oikawa, K. Ishida and T. Kanomata,  Appl. Phys. Lett. 92, 022503 (2008)
\bibitem[17] {17}	Yong-hee Lee, Mitsuharu Todai, Takahiro Okuyama, Takashi Fukuda, Tomoyuki Kakeshita, R. Kainuma, Scri. Mater. 64, 927, (2011)
\bibitem[18] {18}	 M. K. Chattopadhyay, K. Morrison, A. Dupas, V. K. Sharma, L. S. Sharath Chandra, L. F. Cohen, and S. B. Roy, J. of Appl. Physics 111 (2012) 053908 
\bibitem[19] {19}	V. K. Sharma, M. K. Chattopadhyay, and S. B. Roy, Phys. Rev. B 76, 140401(R) (2007)
\bibitem[20] {20}	A. Banerjee, S. Dash, Archana Lakhani, P. Chaddah, X.Chen, R.V. Ramanujan , Sol. Stat Comm. 151, (2011) 971 
\bibitem[21] {21}	Archana Lakhani, S. Dash, A. Banerjee, P. Chaddah, X. Chen, and R. V. Ramanujan, Appl. Phys. Lett. 99, (2011) 242503 
\bibitem[22] {22}	A. Banerjee, P. Chaddah, S. Dash, Kranti Kumar, Archana Lakhani, X. Chen and R. V. Ramanujan  , Phys. Rev. B , 84, 214420 (2011)
\bibitem[23] {23} Kranti Kumar, A. K. Pramanik, A. Banerjee, and P. Chaddah, S. B. Roy, S. Park,  C. L. Zhang, and S.W. Cheong, Phys. Rev. B, 73, (2006) 184435
\bibitem[24] {24} A. Banerjee, A.K. Pramanik, Kranti Kumar and P. Chaddah, J. Phys. Cond. Matt.  18, L605 (2006)
\bibitem[25] {25} P. Chaddah, Kranti Kumar and A. Banerjee, Phys. Rev.  B, 77, 100402(R) (2008) 
\bibitem[26] {26}	A. Banerjee, K. Mukherjee, Kranti Kumar, and P. Chaddah, Phys. Rev. B,  74 (2006), 224445
\bibitem[27] {27}	R. Rawat, K. Mukherjee, K. Kumar, A. Banerjee, and P. Chaddah, J. Phys.: Condens. Matt. 19 (2007) 256211
\bibitem[28] {28}	P. Kushwaha, R. Rawat, and P. Chaddah, J. Phys.: Condens. Matter 20, (2008) 022204
\bibitem[29] {29}	P. Chaddah , Pramana -J. of Physics 67, 113 (2006)
\bibitem[30] {30}	A. Banerjee, R. Rawat, K. Mukherjee and P. Chaddah, Phys. Rev. B 79, (2009) 212403
\bibitem[31] {31}	Y. Imry and M. Wortis, Phys. Rev. B 19, 3580 (1979).
\bibitem[32] {32}	F. Macia, A. Hernandez-Minguez, G. Abril, J. M. Hernandez, A. Garcia-Santiago, J. Tejada, F. Parisi, and P. V. Santos,  Phys. Rev. B 76, 174424 (2007)
\bibitem[33] {33}	Pallavi Kushwaha, Archana Lakhani, R. Rawat, and P. Chaddah  Phys. Rev. B 80, 174413 (2009)
\bibitem[34] {34}	A. Banerjee, Kranti Kumar and P. Chaddah, J. Phys.: Condens. Matter, 21 (2009) 026002 
\bibitem[35] {35}	S. B. Roy and M. K. Chattopadhyay,  Phys. Rev. B 79, 052407 (2009)
\bibitem[36] {36}	A.L Greer, Science, 267 (5206) (1995), 1947.
\bibitem[37] {37}	A. Banerjee, Kranti Kumar and P. Chaddah, Phys.: Condens. Matter 20 (2008) 255245
\bibitem[38] {38}	T. Krenke, X. Moya, S. Aksoy, M. Acet, P. Entel, Ll. Manosa, A. Planes, J. of Magnetism and Mag. Mater. 310 (2007) 2788. 
\bibitem[39] {39}	Thorsten Krenke, Mehmet Acet, Eberhard F. Wassermann, Xavier Moya, Lluís Manosa, and Antoni Planes, Phys. Rev. B 72, 014412 (2005)


\end{thebibliography}
\end{document}